\pdfoutput=1
\documentclass{sigplanconf}
\usepackage[usenames,dvipsnames]{color}
\usepackage{paralist}
\usepackage{xspace}
\usepackage{url}
\usepackage{graphicx}
\usepackage{floatflt}
\usepackage{times}
\usepackage{balance}
\usepackage{spverbatim}
\usepackage{amsmath}
\usepackage{latexsym}
\usepackage{listings}
\usepackage{mathdots}
\usepackage{setspace}
\usepackage{hyperref}
\usepackage{subfigure}

\usepackage[T1]{fontenc}

\usepackage{cleveref}
\crefname{section}{\S\!\!\!\;}{\S\S}
\Crefname{section}{\S}{\S\S}

\usepackage[utf8]{inputenc} 

\usepackage[scaled]{beramono}

\newcommand{\mysinglespacing}{%
  \setstretch{1}
}
\newtheorem{example}{Example}
\newtheorem{query}{Query}
\newcommand{\squeezeup}{\vspace{-2.5mm}}

\lstMakeShortInline[columns=fixed]!

\usepackage{etoolbox}
\makeatletter
\patchcmd{\maketitle}{\@copyrightspace}{}{}{}
\makeatother

\newcommand{\q}[1]{{\vspace{.1in}\noindent \em {\bf Research Question}}}

\makeatletter
\def\@copyrightspace{\relax}
\makeatother

\newcommand{\cut}[1]{}

\newcommand{\eat}[1]{}

\newcommand{\tr}[1]{}
\newcommand{\tleftout}[1]{}

\newcommand{\squishlist}{\begin{list}{$\bullet$}{\leftmargin=1.2em \topsep 1pt \itemsep -1pt \parsep 1pt}}
\newcommand{\squishend}{\end{list}}

\begin{document}
\title{Towards a Unified Query Language for Provenance and Versioning}

\authorinfo{Amit Chavan}
       {University of Maryland}
       {\url{amitc@cs.umd.edu}}
\authorinfo{Silu Huang}
       {University of Illinois (UIUC)}
        {\url{shuang86@illinois.edu}}
\authorinfo{Amol Deshpande}
       {University of Maryland}
        {\url{amol@cs.umd.edu}}
\authorinfo{Aaron J. Elmore}
       {University of Chicago}
        {\url{aelmore@cs.uchicago.edu}}
\authorinfo{Sam Madden}
       {MIT}
        {\url{madden@csail.mit.edu}}
\authorinfo{Aditya Parameswaran}
       {University of Illinois (UIUC)}
        {\url{adityagp@illinois.edu}}

\maketitle

\newcommand{\vquel}{VQuel\xspace}

\begin{abstract}
Organizations and teams collect and acquire data from various sources, such as social interactions, financial transactions, sensor data, and genome sequencers. Different teams in an organization as well as different data scientists within a team are interested in extracting a variety of insights which requires combining and collaboratively analyzing datasets in diverse ways. DataHub is a system that aims to provide robust version control and provenance management for such a scenario. To be truly useful for collaborative data science, one also needs the ability to specify queries and analysis tasks over the versioning and the provenance information in a unified manner. In this paper, we present an initial design of our query language, called VQuel, that aims to support such unified querying over both types of information, as well as the intermediate and final results of analyses. We also discuss some of the key language design and implementation challenges moving forward.
\end{abstract}

\section{Introduction}
\label{sec:intro}

Data science is becoming increasingly essential in all spheres of existence, and there is a need for new data management tools to facilitate and support data science and collaborative data analysis in general. A key and sorely needed functionality is that of keeping track of datasets at various stages of data analysis, enabling the access, retrieval, and modification of specific dataset versions, and sharing and distribution of datasets with other members of the data analysis team. Often in a collaborative data science scenario, there are hundreds or thousands of versions of collected, curated, and derived datasets, at various degrees of structure (fully structured and relational all the way to completely unstructured), each of which can have millions to billions of records within them.

While \texttt{git} and \texttt{svn} have proved tremendously useful for collaborative source code management, they are inadequate for managing datasets for several reasons~\cite{datahubcidr}.  First, they are based on a model of either ``checking out'' the entire repository (git), or keeping two copies of each file in the working directory (svn), which may not be practical when dealing with large collections of large files.  Second, they employ Unix-diff-like differencing semantics when merging changes.  For text files, this means they identify overlapping ranges of edits, and allow changes in non-overlapping regions.  For relational datasets, this merge policy can be both too restrictive (e.g., in a dataset, if two edits both inserted records, there should not be a conflict), and miss conflicts (e.g., a merge should not be allowed if it results in a violation of a primary key constraint.)  Third, the underlying algorithms are not optimized for large files or repositories, and can be painfully slow in such settings.  Finally, as we focus on in this paper, their versioning API is based a notion of files, not structured records, and as such, is not a good fit for a scenario with a mix of structured and unstructured datasets; the versioning API is also not capable of allowing data scientists to reason about data contained within versions and the relationships between the versions in a holistic manner.

To address these challenges, we are building DataHub, a collaborative data analysis platform~\cite{datahubcidr} aimed at being the ``GitHub for structured data''\footnote{GitHub is a widely used Web-based \texttt{git} repository hosting service}.
DataHub enables the compact storage and indexing of versioned datasets~\cite{datahubcidr, vldb-theory}, and at the same time enables the use of popular data analysis tools such as R or Python via Thrift-based APIs~\cite{thrift} and a hosted execution environment.
What DataHub currently lacks is a query language that is capable of capturing all the desired operations necessary to facilitate collaborative data analysis --- at the present moment, most of this querying is done tediously via hand-written scripts.

In this paper, we present our initial proposal for a {\em version-aware query language}, capable of querying dataset versions, dataset provenance (e.g., which datasets a given dataset was derived from), and record-level provenance (if available).
While there has been some work on temporal query languages~\cite{snodgrass1987temporal}, these languages do not apply to our setting since they assume a linear chain of versions --- in our case, we could have an arbitrary branching structure of versions as is common in collaborative data analysis.
Extensions have been proposed to SQL~\cite{korth1989query} to work with nested relational model which allows for relation-valued attributes (as discussed in the next section, we use a nested hierarchical data model); but overall SQL is ill-suited to traversing a graph structure---one of our key requirements, and further, it has a cumbersome aggregation syntax that results in unwieldy queries when comparing across versions~\cite{date1984critique}.
Similarly, while there has been substantial work on query languages for provenance, ranging from adapting SQL~\cite{trio}, Prolog~\cite{marinho2012provmanager,murta2014}, SPARQL~\cite{kim2008provenance,wylot2015} to specialized languages such as QLP~\cite{anand2009, anand2010}, PQL~\cite{holland2008choosing}, ProQL~\cite{karvounarakis2010querying} (~\cite{bowers2012scientific},~\cite{davidson2008provenance} have additional examples), much of this work centers on well-defined workflows and tuple-based provenance rather than collaborative settings where multiple users interact through a derivation graph of versions in an ad hoc manner. Furthermore, query languages are generally tied to a particular method of recording provenance information, e.g., semiring annotations~\cite{green2007}, COMAD~\cite{bowers2008provenance}, etc., and adapting them to other provenance data and storage models is often clunky~\cite{freire2008provenance}. Finally, we note that although our proposed language is different from the aforementioned ones, we might be able to build upon some of their query execution strategies (e.g.,~\cite{wylot2015}) and add user-defined operators to aid in specific analysis tasks (e.g.,~\cite{murta2014}). This is, however, ongoing work and is not the focus of this paper.

To the best of our knowledge, ours is the first query language proposal tailored for an ad hoc derivation graph of versions of structured records. Our proposal draws from constructs introduced in the historical Quel~\cite{stonebraker1976design} and GEM~\cite{zaniolo1983database} languages, neither of which had a temporal component.

To illustrate the features of our query language, we describe an example collaborative data analysis scenario, and then present examples of queries we would like to issue:

\begin{example}
\emph{Genome assembly} of a whole genome sequence dataset is a complex task --- apart from huge computational demands, it is not always known a priori which tools and settings will work best on the available sequence data for an organism~\cite{baker2012novo}.
The process typically involves testing multiple tools, parameters and approaches to produce the best possible assembly for downstream analysis.
The assemblies are evaluated on a host of metrics (e.g., the N50 statistic) and the choice of which assembly is the best one is also not always clear.
One potential sequence of steps might be:
Sequenced reads (FastQ files) $\rightarrow$ Error correction tools (Quake, Sickle, etc.) $\rightarrow$ Input analysis, k-mer calculation (KmerGenie) $\rightarrow$ Assembly tool (SOAPdenovo, ABySS) $\rightarrow$ Assembly analysis and selection (QUAST).

A group of researchers may collaboratively try to analyze this data in various ways, building upon the work done by the others in the team, but also trying out different algorithms or tools. New data is also likely to be ingested at various points, either as updates/corrections to the existing data or as results of additional experiments. As one can imagine, the ad hoc nature of this process and the desire not to lose any intermediate synthesized result means that the researchers will be left with a large number of datasets and analyses, with large overlaps between them and complex derivational dependencies.
Similar collaborative workflows can be seen in many other data science application domains.
\end{example}

Before moving forward, we describe our notion of the term ``version''. For us, a version consists of one or more datasets that are semantically grouped together (in some sense, it is equivalent to the notion of a ``commit'' in \texttt{git/svn}). A version, identified by an ID, is immutable and any update to a version conceptually results in a new version with a different version ID (note that the physical data structures are not necessarily immutable and we would typically not want to copy all the data over, but rather maintain differences~\cite{vldb-theory}). New versions can  also be created through the application of transformation programs to one or more existing versions. The version-level provenance that captures these processes is maintained as a ``version graph'', that we discuss in more detail later.

There is a wide range of queries that may be of interest in such a setting as above. Simple queries include: (a) identifying versions based on the metadata information (e.g., authors); (b) identifying versions that were derived (directly or through a chain of derivations) from a specific outdated version; and (c) finding versions that differ from their predecessor version by a large number of records. More complex queries include: (d) finding versions where the data within satisfies certain aggregation conditions; (e) finding the intersection of a set of versions (representing, e.g., the final synthesized results of different pipelines); and (f) finding versions that contain any records derived from a specific record in a version. We note here that a key challenge that we face is identifying a useful set of queries/tasks and abstracting language features from them, and we hope to engage with a wide variety of users to accomplish that.


These examples illustrate some of the key requirements for a query language, namely the ability to:
\begin{itemize}
\item Traverse the version graph (i.e., version-level provenance information) and query the metadata associated with the versions and the derivation/update edges.
\item Compare several versions to each other in a flexible manner.
\item Run declarative queries over data contained in a version, to the extent allowable by the structure in the data.
\item Query the tuple-level provenance information, when available, in conjunction with the version-level provenance information.
\end{itemize}

\noindent In the rest of this paper, we describe our proposal for a language, called \vquel, that aims to provide these features. We emphasize that \vquel is a work-in-progress; we fully expect our language to evolve with feedback from end-users.

\section{Preliminaries}
\label{sec:background}

Like GitHub, DataHub aims at enabling data scientists host, share, and manage datasets with ease. Unlike GitHub, however, to support end-user tasks directly inside DataHub, the platform also hosts data-processing apps.
This makes DataHub a platform for sharing datasets, along with the computation on those datasets. Details on DataHub are covered in a recent overview paper~\cite{datahubcidr} and at \url{http://datahub.csail.mit.edu/}).

\begin{figure*}[htp]
\centering
\subfigure[]{\includegraphics[width=0.95\columnwidth]{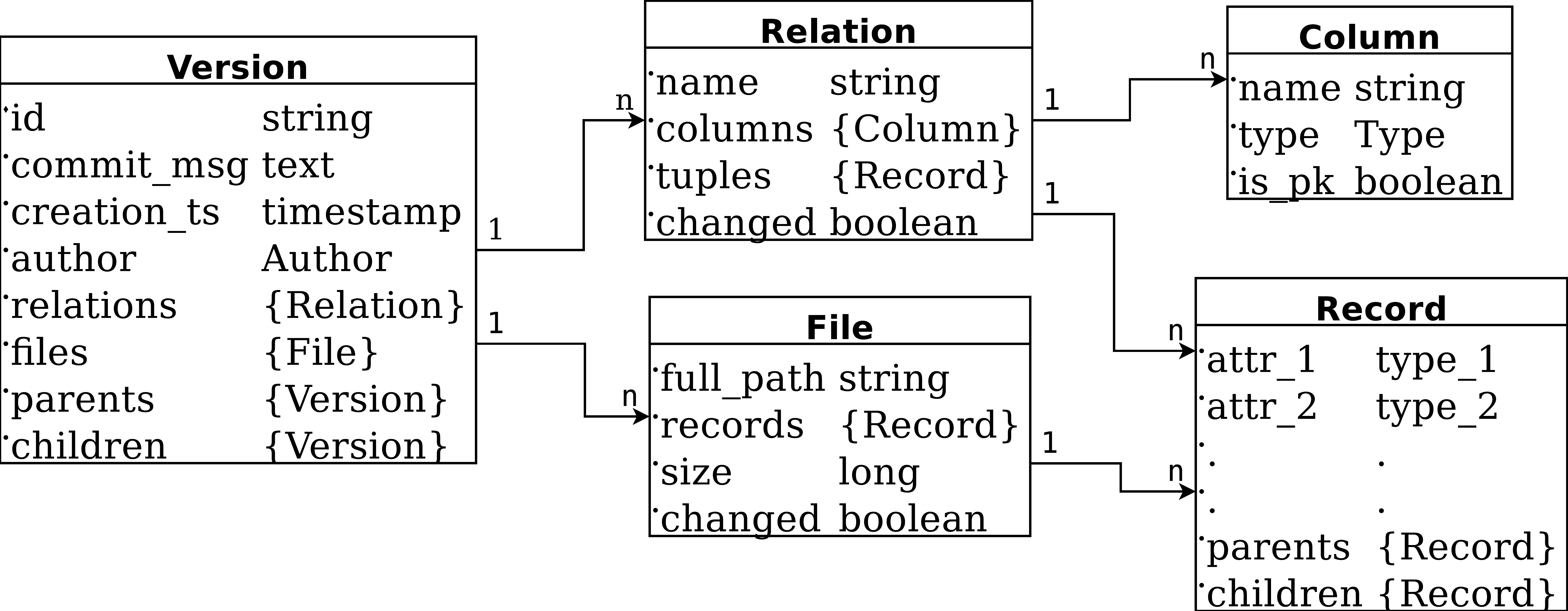}}
\hspace{1cm}
\subfigure[]{\includegraphics[width=1\columnwidth]{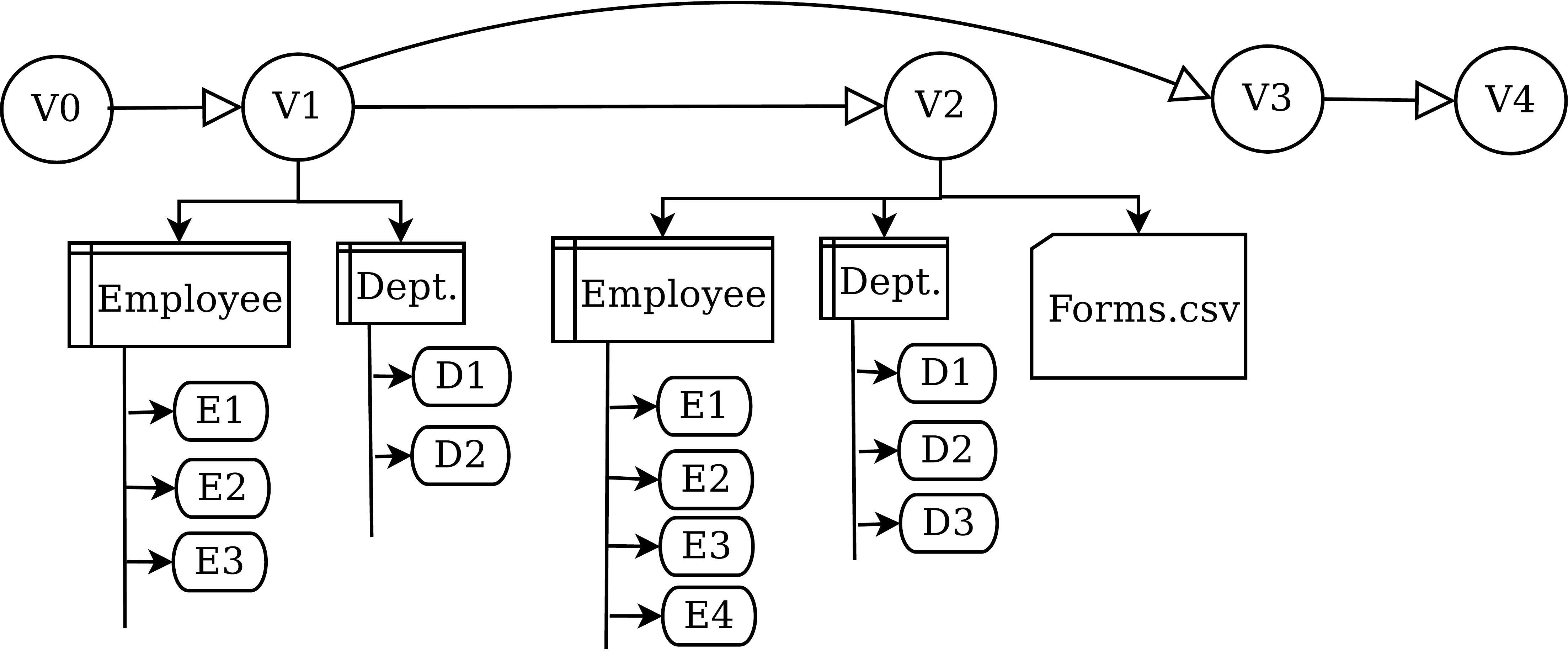}}
\caption{(a) Conceptual Data model of DataHub: the notation ``\{T\}'' denotes a set of values of T; fields in the Records entity can be conceptually thought of as a union of all fields across records; other fields and entities (for instance Authors) are not shown to keep the discussion brief; for each entity, entries in the left and right column denote the attribute name and type respectively. (b) An example version graph where circles denote versions; version V1 has two Relations, {\tt Employee} and {\tt Department}, each having a set of records, \{E1, E2, E3\} and \{D1, D2\} respectively; version V2 adds new records to both the {\tt Employee} and {\tt Department} relations and also adds a new File, \texttt{Forms.csv}. Edge annotations (not shown) are used to capture information about the derivation process itself, including references to transformation programs or scripts if needed.
}
\label{fig:hierarchy}
\end{figure*}

DataHub enables users to keep track of datasets and their versions, by means of a version graph that encodes derivation relationships among them. As we discussed earlier, a version refers to a collection of files or relations that are semantically grouped together. Figure \ref{fig:hierarchy}(b) shows an example of a few versions along with the {\em version graph} connecting them. 

Figure \ref{fig:hierarchy}(a) shows a portion of the conceptual data model that we use to write queries against.
The data model consists of four essential tables:
\texttt{Version}, \texttt{Relation}, \texttt{File}, and \texttt{Record}. Additional tables like \texttt{Column} and \texttt{Author} are required in DataHub but not essential for the purpose of this discussion. The difference between \texttt{Relation} and \texttt{File} is that a relation has a fixed schema for all its records (recorded in the \texttt{Column} table) while a file has no such requirement. To that effect, we denote the records in a relation as tuples.

The \texttt{Version} table maintains the information about the different versions in the database, including the ``commit\_id'' (unique across the versions), and various attributes capturing metadata about the version, such as the creation time and author,
as well as ``commit\_msg'' and ``creation\_ts'', representing the commit message and creation time respectively.
There are four set-valued attributes called ``relations'', ``files'', ``parents'' and ``children'', recording the relations and files contained in the version, and the direct parents and children in version graph respectively. The last two refer back to the \texttt{Version} table, whereas the first two refer to the \texttt{Relation} and {\tt File} tables respectively. A tuple in the  {\tt Relation} table, in turn, records the information for a relation including its schema; we view the tuples in the relation as a set-valued attribute of this table itself --- this allows us to locate a relation and then query on the data inside it as we will see in the next section. The {\tt Files} table is analogous, but records information appropriate for an unstructured file. Note that neither of these tables has a primary key but rather the attributes ``name'' and ``full\_path'' serve as {\em discriminators}, and must be combined with the version ``id'' to construct primary keys. The ``changed'' attribute is a derived (redundant) attribute that indicates whether the relation/file changed from the parent version, and is very useful for version-oriented queries.

Finally, {\tt Record} is a virtual table that can be conceptually thought of as a union of all tuples and records in all relations and files across the versions. The one exception are the ``parents'' and ``children'' attributes, which refer back to the {\tt Record} table and can be used to refer to fine-grained provenance information within queries. This table is never directly referenced in the queries, but is depicted here for completeness. The provenance information must ``obey'' the version graph, e.g., in the example shown, records in version V2 can only have records in version V1 as parents.

We note here that this data model is a high-level conceptual one mainly intended for ease of querying and aims to maximize data independence. For instance, although the fine-grained provenance information is conceptually maintained in the {\tt Record} table here and can be queried using the ``parents'' and ``children'' attributes, the implementation could maintain that information at schema-level wherever feasible to minimize the storage requirements.

\section{Overview of \vquel}
\label{sec:language}
\vquel is largely a generalization of the Quel language (while also introducing certain syntactic conveniences that Quel does not possess), and combines features from GEM and path-based query languages. This means that \vquel is a {\em full-fledged relational query language}, and in addition, it enables the seamless querying of the nested data model described in the previous section, encoding versioning derivation relationships, as well as versioning metadata.


\vquel will be illustrated using example queries on the repository shown in Figure~\ref{fig:hierarchy}(b), with certain deviations introduced when necessary. We will introduce the constructs in \vquel incrementally, starting from those present in Quel to the new ones designed for the DataHub setting. For ease of understanding, we first present a version that is clear and easy to understand, but results in longer queries. In Section~\ref{sec:short} we describe additional constructs to make the queries concise. 

\subsection{Examples}

We begin with some simple \vquel queries.
Most of these queries are also straightforward to write in SQL;
the queries that cannot be written in SQL easily begin in Section~\ref{sec:agg-op}.
Here, we gradually introduce the constructs of \vquel as a prelude to the more complex queries combining versioning and data.

\newpage
\begin{query}
\label{q:find-author}
Who is the author of version with id ``v01''?
\end{query}
\squeezeup
\begin{lstlisting}
range of V is Version
retrieve V.author.name
where V.id = "v01"
\end{lstlisting}

\noindent A \vquel query has two elements: iterator setup (!range! above) and retrieval (!retrieve! above) of objects satisfying a predicate (!where! above). Iterators in \vquel are similar to tuple variables in Quel, but more powerful, in the sense that they can iterate over objects at any level of our hierarchical data model. They are declared with a statement of the form:
\begin{lstlisting}
range of <iterator-variable> is <set>
\end{lstlisting}

\noindent The !retrieve! statement is used to select the object properties, and is of the form:
\begin{lstlisting}
retrieve [into <iterator>][unique]<target-list>
[where <predicate>]
[sort by <attribute> [asc/desc] {, <attribute> [asc/desc]}]
\end{lstlisting}

\noindent The !retrieve! statement fetches all the object attributes specified in the \texttt{target-list} for those objects satisfying the !where! clause.

\begin{query}
\label{q:find-commits}
What commits did Alice make after January 01, 2015?
\end{query}
\squeezeup
\begin{lstlisting}
range of V is Version
retrieve V.all
where V.author.name = "Alice" and V.creation_ts >= "01/01/2015"
\end{lstlisting}

\noindent
In Queries~\ref{q:find-author} and \ref{q:find-commits}, note the use of GEM-style tuple-reference attributes, namely !V.author!, and the keyword !all! from Quel. The comparators \lstinline:=, !=, <, <=, >: and \lstinline:>=: are allowed in comparisons, and the logical connectives !and!, !or!, and !not! can be used to combine comparisons.

Multiple iterators can be set up before a retrieval statement, and their respective sets can be defined as a function of previously declared iterators. The next example illustrates this idea.  The first range clause sets up an iterator !V! over all the versions. The second range clause defines an iterator over all relations inside a version.

\begin{query}
List the commit timestamps of versions that contain the Employee relation.
\end{query}
\squeezeup
\begin{lstlisting}
range of V is Version
range of R is V.Relations
retrieve V.commit_ts
where R.name = "Employee"
\end{lstlisting}

\begin{query}
Show the commit history of the Employee relation in reverse chronological order.
\end{query}
\squeezeup
\begin{lstlisting}
range of V is Version
range of R is V.Relations
retrieve V.creation_ts, V.author.name, V.commit_message
where R.name = "Employee" and R.changed = true
sort by V.creation_ts desc
\end{lstlisting}

\noindent Similarly, we can set up a range clause over tuples inside a relation. Analogous to a relational database, the user needs to be familiar with the schema to be able to pose such a query. 

\begin{query}
Show the history of the tuple with employee id ``e01'' from Employee relation.
\end{query}
\squeezeup
\begin{lstlisting}
range of V is Version
range of R is V.Relations
range of E is R.Tuples
retrieve E.all, V.commit_id, V.creation_ts
where E.employee_id = "e01" and R.name = "Employee"
sort by V.creation_ts
\end{lstlisting}

\subsection{Syntactic sweetenings}
\label{sec:short}
In this section, we introduce some shorthand constructs to keep the size of the queries small. These constructs are meant only for brevity, and each of them can be mapped to an equivalent query without using shorthands.

The first one is analogous to a {\em filter} operation over a set declaration: we can use predicates in the set declaration block of the !range! statement. For instance, in the following example, both queries iterate over the same set of versions. Note that the !retrieve into! clause in (b1) sets up a new iterator !V! over all the versions satisfying constraints in !where! clause.

\begin{lstlisting}
(a1) range of V is Version(id = "v01")

(b1) range of T is Version
     retrieve into V (T.all)
     where T.id = "v01"
\end{lstlisting}

\noindent The next example shows the principle in action on a query that would otherwise become quite long. Again, (a2) and (b2) below show identical queries written using the short notation (a) and their equivalent form (b).

\begin{query}
Find all Employee tuples in version ``v01'' that are different in version ``v02''.
\end{query}
\squeezeup
\begin{lstlisting}
(a2) range of E1 is Version(id = "v01").Relations(name = "Employee").Tuples
     range of E2 is Version(id = "v02").Relations(name = "Employee").Tuples
     retrieve E1.all
     where E1.employee_id = E2.employee_id and E1.all != E2.all

(b2) range of V1 is Version
     range of R1 is V1.Relations
     range of E1 is R1.Tuples
     range of V2 is Version
     range of R2 is V2.Relations
     range of E2 is R2.Tuples
     retrieve E1.all
     where V1.id="v01" and R1.name="Employee"
     and V2.id="v02" and R2.name="Employee"
     and E1.employee_id = E2.employee_id and E1.all != E2.all
\end{lstlisting}

\subsection{Aggregate operators}\label{sec:agg-op}
The aggregate functions !sum!, !avg!, !count!, !any!, !min! and !max! are also provided in \vquel. Any expression involving components of iterated entity attributes, constants and arithmetic symbols can be used as the argument of these functions. Due to the nested nature of iterators, we introduce the~!_all! version of these operators, i.e. !count_all!, !sum_all!, etc. The general syntax of an aggregate expression is:
\begin{lstlisting}
agg_op([<agg-attribute>/<iterator-variable>] [group by <grouping-attributes>] [where <predicate>])
\end{lstlisting}

This evaluates the !agg_op! on each group of !<agg-attribute>! of objects that satisfy the !<predicate>!. We see two examples next.

\begin{query}
\label{q:simple-agg-1}
For each version, count the number of relations inside it.
\end{query}
\squeezeup
\begin{lstlisting}
range of V is Version
range of R is V.Relations
retrieve V.id, count(R)
\end{lstlisting}

\begin{query}
\label{q:simple-agg-2}
Find all versions containing precisely 100 Employees with last name ``Smith''.
\end{query}
\squeezeup
\begin{lstlisting}
range of V is Version
range of E is V.Relations(name = "Employee").Tuples
retrieve V.commit_id
where count(E.employee_id where E.last_name = "Smith") = 100
\end{lstlisting}

\noindent In both queries above, the aggregation is performed only over objects at the \emph{innermost} level of an iterator expression. In query~\ref{q:simple-agg-1}, !R! is an iterator over relations inside a version !V!, and !count! iterates only over the innermost level of this iterator hierarchy, that is, !R!. Similarly, in query~\ref{q:simple-agg-2}, the !count! expression only iterates over the tuples inside a relation inside a version.

Notice that the latter query is not very easy to express in vanilla SQL:
there is no easy way to use SQL to retrieve version numbers, which in a traditional non-versioned context would either be considered as schema-level information, or involve multiple joins depending on the level of normalization of the schema.
 \vquel, on the other hand, allows us to set up the nested iterators that makes such queries very easy to express.

The next two examples show the usage of !count_all! operator. The difference from the !count! operator is that all the ``parent'' iterators are evaluated, instead of only the innermost iterator, to compute the value of the aggregate. Another way to reason about this behavior is that !count! has an implicit grouping list of attributes in its by clause: query~\ref{q:simple-agg-all} is identical to query~\ref{q:simple-agg-2}.

\begin{query}
\label{q:simple-agg-all}
Find all versions containing precisely 100 employees with last name ``Smith''.
\end{query}
\squeezeup
\begin{lstlisting}
range of V is Version
range of R is V.Relations(name = "Employee")
range of E is R.Tuples
retrieve V.commit_id
where count_all(E.employee_id group by R, V where E.last_name = "Smith") = 100
\end{lstlisting}

\noindent Aggregates having a !group by! clause can also be used in the predicate to restrict the results of the query. In query~\ref{q:simple-agg-all}, the result of !count_all! for each group is compared against !100!. Query~\ref{q:filter-by-agg} gives another example.

\begin{query}
\label{q:filter-by-agg}
Find all versions containing precisely 100 tuples in all relations put together inside a version.
\end{query}
\squeezeup
\begin{lstlisting}
range of V is Version
range of R is V.Relations
range of T is R.Tuples
retrieve V.all
where count_all(T group by V) = 100
\end{lstlisting}

\noindent The next few examples show how we can use aggregate operators across a set of versions to answer a variety of questions about the data.


\begin{query}
Among a group of versions, find the version containing most tuples that satisfy a predicate. For instance, which version contains the most number of employees above age 50?
\end{query}
\squeezeup
\begin{lstlisting}
range of V is Version
range of E is V.Relations(name = "Employee").Tuples
retrieve into T (V.id as id, count(E.id where E.age > 50) as c)
retrieve T.id
where T.c = max(T.c)
\end{lstlisting}

\noindent Up until now, for an iterator, we have been exploring ``down'' the hierarchy. We also provide appropriate functions, depending on the type of iterator, to refer to values of entities ``up'' in the hierarchy. In the next query, !Version(T)! is used to refer to the version attributes of tuples in !T!.

\begin{query}
Which versions are such that the natural join between relations S and T has more than 100 tuples?
\end{query}
\squeezeup
\begin{lstlisting}
range of V is Version
range of S is V.Relations(name = "S").Tuples
range of T is V.Relations(name = "T").Tuples
retrieve into Q(V.id as id,
    count_all(S.id group by V where S.id = T.s_id and Version(S).id = Version(T).id) as c)
retrieve Q.id
where Q.c >= 100
\end{lstlisting}

\subsection{Version graph traversal}
\vquel has three constructs aimed at traversing the version graph. Each of these operate on a version at a time, specified over an iterator.
\begin{itemize}
\item
!P(<integer>)!: Return the set of ancestor version of this version, until integer number of hops in the version graph. If the number of hops is not specified, we go till the first version. Duplicates are removed.
\item
!D(<integer>)!: Similar to !P()! except that it returns the descendant/derived versions.
\item
!N(<integer>)!: Similar to !P()! except that it returns the versions that are !<integer>! number of hops away.
\end{itemize}

\noindent The next few queries illustrate these constructs. Notice once again that queries of this type are not very easy to express in SQL, which does not permit the easy traversal of graphs, or specification of path queries. The constructs we introduce are reminiscent of constructs in graph traversal languages~\cite{woodgraph}; these combined with the rest of the power of \vquel enable some fairly challenging queries to be expressed rather easily.

\begin{query}
Find all versions within 2 commits of ``v01'' which have less than 100 employees.
\end{query}
\squeezeup
\begin{lstlisting}
range of V is Version(id = "v01")
range of N is V.N(2)
range of E is N.Relations(name = "Employee").Tuples
retrieve N.all
where count(E) < 100
\end{lstlisting}

\begin{query}
Find all versions where the delta from the previous version is greater than 100 tuples.
\end{query}
\squeezeup
\begin{lstlisting}
range of V is Version
range of P is V.P(1)
retrieve unique V.all
where abs(count(V.Relations.Tuples) - count(P.Relations.Tuples)) > 100
\end{lstlisting}

\begin{query}
For each tuple in Employee relation as of version ``v01'', find the parent version where it first appeared.
\end{query}
\squeezeup
\begin{lstlisting}
range of V is Version(id = "v01")
range of E is V.Relations(name = "Employee").Tuples
range of P is V.P()
range of PE is P.Relations(name = "Employee").Tuples
retrieve E.id, P.id
where E.employee_id = PE.employee_id and P.commit_ts = min(P.commit_ts)
\end{lstlisting}

\subsection{Extensions to fine-grained provenance}
\label{sec:provenance}

Finally, in some cases, we may have complete transparency into the operations performed by data scientists. In such cases, we can record, reason about, and access tuple-level provenance information. Here is an example of a query that can refer to tuple-level provenance:

\begin{query}
For tuples in version ``v01'' in relation S that satisfy a predicate, say value of attribute !attr = x!, find all parent tuples that they depend on.
\end{query}
\squeezeup
\begin{lstlisting}
range of E is Version(id = ``v01'').Relations(name = ``S'').Tuples
range of P is E.parents
retrieve E.id, P.id
where E.attr = x
\end{lstlisting}

\noindent Similar queries can be used to ``walk up'' the derivation path of given tuples, for example, to identify the origins of specific tuples.

\section{Implementation Challenges}
\label{sec:challenges}

In addition to designing our high-level query language, \vquel, we are, in parallel, building an execution and optimization engine for efficiently processing such \vquel queries over large volumes of versioned datasets. A key challenge here is that the data must be stored in a compressed fashion, by exploiting the overlaps across versions. In our recent work~\cite{vldb-theory}, we formally analyzed the tradeoff between the storage needs and the recreation cost of reconstructing specific versions. At one extreme, storing all the versions independently of each other results in the lowest recreation cost but with prohibitively high storage costs; on the other hand, trying to optimize storage on its own by attempting to exploit the overlaps fully, typically leads to unacceptably high recreation costs. In our work, we presented algorithms to balance these two objectives in a principled manner. In an independent line of work, we are also investigating different options to maintain a large number of versions of a relational database in a concise manner within the database through customized storage engines.

A natural way to build upon that work for the query language described here would be to: (a) execute a portion of the query on the metadata and the version graph to identify the versions referenced in the query, and (b) recreate those versions in their entirety, and (c) execute the remainder of the query on those versions, via iterative execution as directly specified in the query, without any query rewriting. Even with this baseline approach, there are several challenges that need to be addressed. For example, the version graph could grow over time to be fairly large, especially in highly dynamic environments where data is continuously being ingested or analyzed (and resulting versions stored). It is also not obvious how to execute some of the comparison functionality where the user wants to reason about differences between large versions.

However a bigger implementation challenge is to develop query execution techniques that can work directly on the compressed representations. Consider, for instance, a query asking for the version with the highest value of an aggregate. The naive approach would be to compute the aggregate for each version separately, which would likely have very high computational cost for reasonable numbers of versions. If the versions are largely overlapping (i.e., the differences between them are small), we must be able to share the computation across them. Doing this in a principled fashion for different types of storage mechanisms is a major research challenge that we are presently pursuing.


{\scriptsize
\bibliographystyle{abbrv}
\bibliography{ref}
}
\end{document}